\documentclass[aps,reprint,prl,showpacs,superscriptaddress,floatfix]{revtex4-1}
\usepackage{amssymb,amsmath,amsfonts,amstext}
\usepackage{amsbsy}
\usepackage{txfonts,color}

\usepackage{epsfig,graphicx}
\usepackage{hyperref}
\hypersetup{colorlinks,citecolor=NavyBlue,filecolor=black,linkcolor=black,urlcolor=black}
\usepackage{color}
\usepackage[dvipsnames]{xcolor}

\usepackage{multibib}

\definecolor{blueRP}{rgb}{0.0, 0.58, 0.71}

\newcommand{\adaga}{a^{\dagger}a}
\newcommand\ket[1]{\left|#1\right\rangle}
\newcommand\bra[1]{\left\langle #1 \right|}

\newcommand\etaIn{\eta\rightarrow\infty}

\begin{document}
\date{\today}
\title[]{Universal anti-Kibble-Zurek scaling in fully-connected systems}
\author{Ricardo Puebla}
\affiliation{Centre for Theoretical Atomic, Molecular, and Optical Physics, School of Mathematics and Physics, Queen's University Belfast, Belfast BT7 1NN, United Kingdom}
\author{Andrea Smirne}
\affiliation{Institute of Theoretical Physics and IQST, Albert-Einstein Allee 11, Universit\"{a}t Ulm, 89069 Ulm, Germany}
\affiliation{Dipartimento di Fisica ''Aldo Pontremoli'', Universit\`a degli Studi di Milano, e Istituto Nazionale di Fisica Nucleare, Sezione di Milano, Via Celoria 16, I-20133 Milan, Italy} 
\author{Susana F. Huelga}
\affiliation{Institute of Theoretical Physics and IQST, Albert-Einstein Allee 11, Universit\"{a}t Ulm, 89069 Ulm, Germany}
\author{Martin B. Plenio}
\affiliation{Institute of Theoretical Physics and IQST, Albert-Einstein Allee 11, Universit\"{a}t Ulm, 89069 Ulm, Germany}
\begin{abstract}
  We investigate the quench dynamics of an open quantum system involving a quantum phase transition. In the isolated case, the quench dynamics involving the phase transition exhibits a number of scaling relations with the quench rate as predicted by the celebrated Kibble-Zurek mechanism.  In contact with an environment however, these scaling laws breakdown and one may observe an anti-Kibble-Zurek behavior: slower ramps lead to less adiabatic dynamics, increasing thus non-adiabatic effects with the quench time. In contrast to previous works, we show here that such anti-Kibble-Zurek scaling can acquire a universal form in the sense that it is determined by the equilibrium critical exponents of the phase transition, provided the excited states of the system exhibit singular behavior, as observed in fully-connected models. This demonstrates novel universal scaling laws granted by a system-environment interaction in a critical system.  We illustrate these findings in two fully-connected models, namely, the quantum Rabi and the Lipkin-Meshkov-Glick models. In addition, we discuss the impact of non-linear ramps and finite-size systems. 
\end{abstract}
\maketitle

{\em Introduction.---} The scrutiny of quantum matter driven out of equilibrium has led to the discovery of novel and striking phenomena~\cite{Eisert:15,Polkovnikov:11}. A comprehensive understanding of out-of-equilibrium properties of quantum systems is of crucial relevance for the further development of quantum technologies, as for example the exploitation of adiabatic evolution for quantum state preparation and computation~\cite{Cirac:12,Albash:18} or for the design and benchmark of quantum simulators~\cite{Buluta:09,Georgescu:14}. In this regard, a remarkable aspect of many-body quantum systems consists in the existence of quantum phase transitions (QPT)~\cite{Sachdev}, which entails a sudden change in their ground state at a critical value of a control parameter. The critical point where a QPT takes place is typically accompanied by a vanishing energy gap~\cite{Sachdev}, thus challenging the success of an adiabatic driving across it.

The Kibble-Zurek (KZ) mechanism, originally proposed to account for defect formation across a symmetry-breaking phase transition in the early universe, has become a cornerstone in non-equilibrium critical dynamics~\cite{Zurek:96,Laguna:97,Zurek:13,delCampo:14} and on the universal behavior of quench dynamics~\cite{Kolodrubetz:12,Acevedo:14,Puebla:17c,Puebla:17,Puebla:19}. Its key prediction consists in a  scaling relation between the number of defects formed upon traversing a phase transition, the equilibrium critical exponents and the quench rate. Such scaling relation is {\em universal} as it is solely determined by the equilibrium critical exponents of the phase transition, and hence, KZ predictions hold in different systems as confirmed in~\cite{Pyka:13,Ulm:13,Monaco:02,Navon:15,Yukalov:15,Beugnon:17}. 
Indeed, the KZ mechanism also applies to the quantum realm~\cite{Zurek:05,Damski:05,Dziarmaga:05,Polkovnikov:05}, where scaling relations are found also for quantum excitations produced during the quench towards or across a QPT~\cite{Dziarmaga:10,Clark:16,Anquez:16,Silvi:16,Xu:14,Cui:16,Gong:16,Keesling:19,Puebla:19}. Remarkably, although in these settings an adiabatic evolution is hindered by a vanishing energy gap, the scaling laws dictated by the KZ mechanism still imply a smaller number of excitations for slower quenches. By reducing the quench rate, the adiabatic condition is eventually achieved, although with a distinctive and smaller scaling exponent than in non-critical systems~\cite{deGrandi:10b}.

The significant experimental progress in the last decades has enabled an unprecedented degree of control, manipulation and preparation in quantum many-body systems~\cite{Friedenauer:08,Kim:09,Kim:10,Islam:11,Islam:13,Richerme:14,Jurcevic:14,Cohen:15,Smith:16,Jurcevic:17}, opening the door for the realization of quantum simulators and computers.  However, the isolation from any environmental disturbance and/or experimental imperfection remains a formidable challenge.
In this regard, it is worth mentioning that an interaction between the system of interest and its surroundings can have a dramatic impact in the properties of the system even when they interact weakly~\cite{Breuer,Rivas}, as demonstrated by the novel phenomena taking place in different dissipative critical systems~\cite{DallaTorre:10,Nagy:15,Yin:16,Hedvall:17,Hwang:18,Kirton:19,Nigro:19,Rossini:19}.  It is therefore important to study the properties of the KZ mechanism in the presence of an environment. 
As observed in recent studies~\cite{Griffin:12,Nigmatullin:16,Patane:08,Patane:09,Yin:14,Nalbach:15,Dutta:16,Gao:17}, the open nature of the dynamics leads to a departure from the KZ scaling prediction for the isolated case. These observations are encompassed under the term anti-Kibble-Zurek (AKZ) behavior, which refers to a  linear increase of the number of excitations with the quench time. These results suggested that the AKZ behavior looses its universal fingerprints, i.e., the scaling laws as a function of the quench time no longer depend on the equilibrium critical exponents.

Here we show that in certain systems the AKZ behavior itself can acquire a universal form, and is thus in general different from a linear scaling. A driven open quantum system undergoing a QPT can show  power-law relations as a function of the quench time, whose scaling is determined solely by its equilibrium critical exponents as for isolated KZ scaling laws. Such universal AKZ relation crucially depends on the critical behavior of the excited states. We illustrate our findings in two fully-connected  critical systems, namely, the quantum Rabi model (QRM)~\cite{Rabi:36} and the Lipkin-Meshkov-Glick (LMG) model~\cite{Lipkin:65}, taking into account the interaction with an environment. Further, we investigate the impact of non-linear ramps and finite-size effects. Our results indicate novel universal scaling laws emerging in the quench dynamics of an open quantum system involving a QPT.

\begin{figure}
\centering
\includegraphics[width=1\linewidth,angle=0]{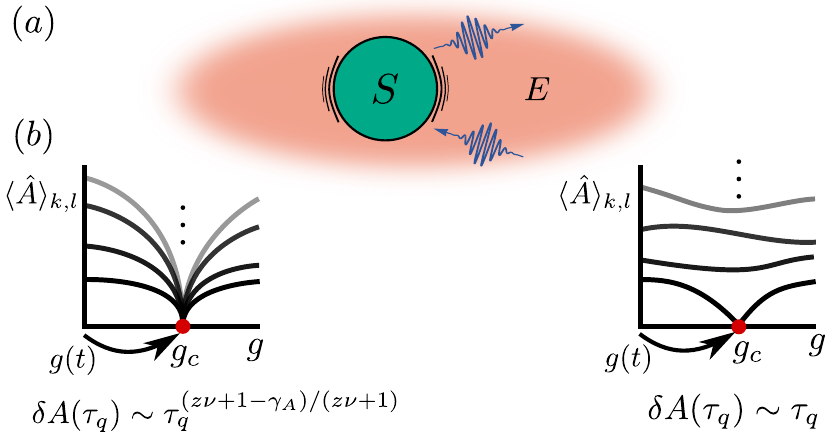}
\caption{\small{(a) Schematic representation of the system $S$ interacting with an environment $E$. In (b) the behavior of the excited states: at the left the excited states behave as the ground state close to $g_c$, thus featuring critical behavior $\langle \hat{A}\rangle_{k,l}\sim |g-g_c|^{\gamma_A}$ for a certain observable $\hat{A}$ whose critical exponent is $\gamma_A$. Then, the excess in $\langle \hat{A}(\tau_q)\rangle_{\rm op}$ with respect to the isolated case, i.e. $\delta A(\tau_q)$, acquires a universal AKZ scaling. For non-critical excited states, the excess $\delta A(\tau_q)$ obeys a linear scaling (see main text for details).}}
\label{fig1}
\end{figure}

{\em Kibble-Zurek mechanism.---} Let us denote by $\hat{H}_S(t)$ the time-dependent  Hamiltonian of an isolated system which drives an initially-prepared ground state across or to the critical point $g_c$ of a QPT by tuning a control parameter $g(t)$ in a total quench time $\tau_q$. Due to the QPT, the energy difference between the first-excited and ground state vanishes at $g_c$ as $\Delta_1(g)\sim |g-g_c|^{z\nu}$~\cite{Sachdev}, where $z$ and $\nu$ denote the dynamic and correlation-length critical exponents of the QPT. This sets a time scale, $\tau_r(g)=\Delta_1^{-1}(g)$, which diverges at $g_c$ thus impeding adiabatic dynamics for a finite quench time and eventually leading to excitations depending on $\tau_q$, as dictated by the KZ mechanism.

The KZ mechanism is built upon the adiabatic-impulse approximation, which relies on the competition between two timescales, namely $\tau_r(g)=\Delta_1^{-1}(g)$ and $t_r(g)=\Delta_1(g)/\dot{\Delta}_1(g)$, where the latter determines the timescale on which the external parameter changes~\cite{Zurek:05,deGrandi:10b,delCampo:14}. For a linear quench $g(t)\propto t/\tau_q$ with $g(0)<g_c$, one finds $t_r(g)\propto \tau_q|g-g_c|$. Within this simplified picture the evolution is split in two regimes as $g(t)$ approaches $g_c$: when $t_r(g)> \tau_r(g)$ the dynamics is fully adiabatic, while the impulse regime is found for $t_r(g)\lesssim \tau_r(g)$. In the latter regime the state {\em freezes} due to the lack of time to adjust to the externally-imposed $g(t)$. Since $\tau_r(g_c)\rightarrow \infty$, the state will eventually cease to follow the ground state of $\hat{H}_S(t)$ close to $g_c$. Hence, the population of excited states and relevant quantities after the quench will depend on $\tau_q$. This heuristic argument is extremely useful to derive the scaling relations in the quench dynamics~\footnote{We note that these arguments can be cast rigorously in a mathematical form that allows for the treatment of general situations~\cite{Gor:16}}. In particular, the boundary between the adiabatic and impulse regime takes place at $\tilde{g}$ such that $t_r(\tilde{g})=\tau_r(\tilde{g})$, which leads to $|\tilde{g}-g_c|\sim \tau_q^{-1/(z\nu+1)}$. Provided  $\tau_q$ is sufficiently long such that diabatic excitations occur due to the QPT, the number of excitations, defined as $n_{\rm ex}=\sum_{k>0} |c_k|^2$, will scale as $n_{\rm ex}\sim \tau_q^{-d\nu/(z\nu+1)}$, where $|\psi(\tau_q)\rangle=\sum_{k=0}c_k \left| \phi_k(g(\tau_q))\right.\rangle$ is the final state and $\hat{H}_S(t)=\sum_{k=0} \epsilon_k\ket{\phi_k(g(t))}\bra{\phi_k(g(t))}$, for a system with $d$ spatial dimensions. In general, an observable $\hat{A}$  whose ground-state expectation value follows $\langle \hat{A}\rangle_0\sim |g-g_c|^{\gamma_A}$ close to $g_c$ being $\gamma_A$ its associated critical exponent, will display KZ scaling according to $\langle \psi(\tau_q)|\hat{A} |\psi(\tau_q)\rangle\sim \tau_q^{-(d\nu+\gamma_A)/(z\nu+1)}$ if $g(\tau_q)=g_c$ or $\tau_q^{-d\nu/(z\nu+1)}$ for $g(\tau_q)> g_c$~\cite{Deng:08,deGrandi:10b,deGrandi:10,deGrandi:10c}. These universal scaling relations are the key KZ predictions for isolated systems. In the following we consider fully-connected models, i.e. $d=0$, so that KZ scaling appears only when $g(\tau_q)=g_c$~\cite{Hwang:15,Defenu:18}.

{\em Universal anti-Kibble-Zurek scaling.---} Let us consider now the open quantum system dynamics and assume in particular a weak interaction with a Markovian environment, such that $\dot{\hat{\rho}}=-i[\hat{H}_S(t),\hat{\rho}]+\mathcal{D}[\hat{\rho}]$, where $\mathcal{D}[\bullet]$ accounts for the dissipative dynamics via a proper (Lindblad) structure, with an overall rate $\kappa$ which expresses the strength of the system-environment interaction~\cite{Breuer,Rivas} (see Fig~\ref{fig1}(a))~\footnote{This can be obtained, for example, if the system interacts with a bosonic environment characterized by the Cauchy-Lorentzian spectral density $I(\omega')= \kappa \lambda^2/((\omega'-\omega)^2+\lambda^2)$, with $\omega$ a reference frequency resonant with the system, in the regime $\omega \gg \lambda \gg \kappa$; see~\cite{Amato:19}.}. In the weak coupling limit and for a finite quench time, the expectation values of the resulting open-system observables (denoted with subscript ${\rm op}$), can be split in two contributions, 
  \begin{align}\label{eq:Aopt}
\langle \hat{A}(\tau_q)\rangle_{\rm op}\equiv {\rm Tr}[\hat{\rho}(\tau_q)\hat{A}]= \langle \hat{A}(\tau_q)\rangle+\delta A(\tau_q),
  \end{align}
  where $\langle \hat{A}(\tau_q)\rangle\equiv \langle \psi(\tau_q)|\hat{A} |\psi(\tau_q)\rangle$ is the isolated (fully coherent) contribution, while $\delta A(\tau_q)$ accounts for the excess introduced by the dissipative dynamics~\cite{Patane:08,Patane:09,Dutta:16,Gao:17}. As aforementioned, resorting to KZ arguments one obtains scaling predictions for $\langle \hat{A}(\tau_q)\rangle$ in terms of $\gamma_A$, $d$, $z$ and $\nu$. On the other hand, $\delta A(\tau_q)$ stems from the contact with the environment and it produces excitations at a constant rate $\kappa$ per unit of time,
  \begin{align}\label{eq:deltaA}
\delta A(\tau_q)\approx \kappa \tau_q \sum_{k,l=0}h_{k,l} \langle\hat{A}\rangle_{k,l}, 
  \end{align}
  for sufficiently small $\kappa\tau_q$, and where $h_{k,l}$ takes account of how the dissipative dynamics populate different excited states and their coherences, while $\langle\hat{A}\rangle_{k,l}\equiv \langle \phi_k(g_f)|\hat{A} |\phi_l(g_f)\rangle $ with $g_f\equiv g(\tau_q)$; see~\cite{sup} for details. If the excited states show critical behavior, that is, if $\langle \hat{A}\rangle_{k,l}\sim |g-g_c|^{\gamma_A}$, assuming the same critical exponent $\gamma_A$  for any $k,l$, then $\delta A(\tau_q)\sim \kappa \tau_q |g_f-g_c|^{\gamma_A}$. Note that this is the case for fully-connected systems~\cite{Caprio:08,Ribeiro:07,Ribeiro:08,Hwang:15,Puebla:16,Larson:17}. Finally, relying on the adiabatic-impulse approximation and introducing $|\tilde{g}-g_c|\sim \tau_q^{-1/(z\nu+1)}$ one finds
  \begin{align}\label{eq:deltaAKZ}
\delta A(\tau_q)\sim \tau_q^{(z\nu+1-\gamma_A)/(z\nu+1)}.
    \end{align}
  This is the key result of the paper. The contribution due to the dissipative dynamics introduces a universal AKZ scaling, as its value is given by the equilibrium critical exponents of the QPT. In contrast, if the excited states do not show critical features (cf. Fig~\ref{fig1}(b)), Eq.~\eqref{eq:deltaA} yields a linear scaling $\delta A(\tau_q)\sim \tau_q$, as for a one-dimensional transverse-field Ising model, which agrees with previous observations~\cite{Patane:08,Patane:09,Nalbach:15,Dutta:16,Gao:17}.  
In~\cite{sup} one can find further details 
about the derivation of Eq.~\eqref{eq:deltaAKZ}, the proof that it can hold even when only few excited states are critical, 
and the analysis of the scaling of the optimal quench time that minimizes Eq.~\eqref{eq:Aopt} with the rate $\kappa$~\cite{Dutta:16}.
In the following, we show the validity of Eq.~\eqref{eq:deltaAKZ} in some case studies.

\begin{figure}
\centering
\includegraphics[width=1.\linewidth,angle=0]{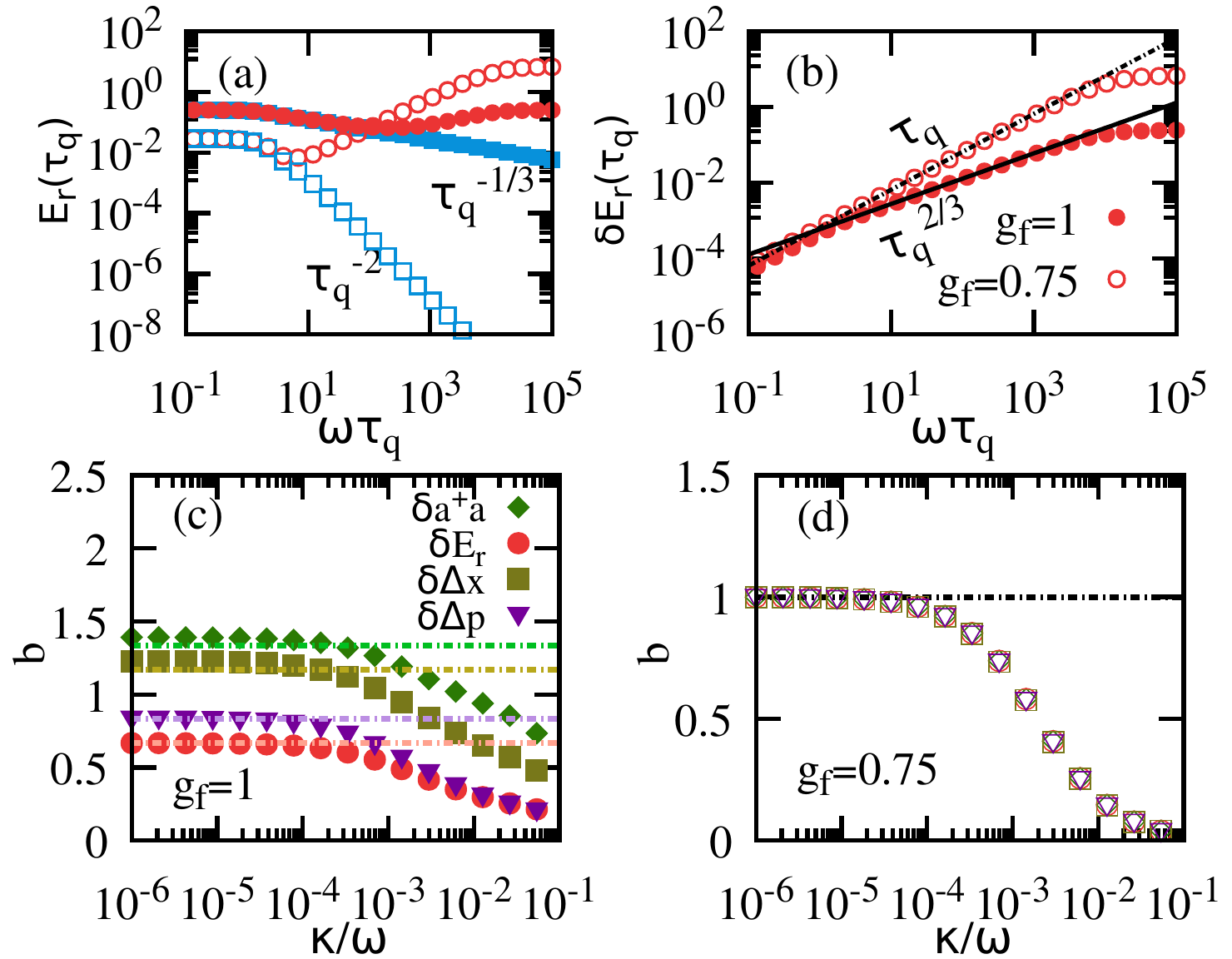}
\caption{\small{(a) Residual energy (in units of $\omega$) for the isolated case $\kappa=0$, $\langle E_r(\tau_q)\rangle$ (blue squares), and for $\kappa=10^{-4}\omega$ and $T=10\omega$, $\langle E_r(\tau_q)\rangle_{\rm op}$ (red circles) for $g_f=1$ (full points) and $g_f=0.75$ (open points) as a function of $\tau_q$.  In (b) we show the excess $\delta E_r(\tau_q)$ for the two cases, unveiling the linear AKZ scaling for $g_f=0.75$ (dashed line) and its universal form when $g_f=1$,  $\delta E_r(\tau_q)\sim\tau_q^{2/3}$ (solid line).  In panel (c) we show the fitted exponent $b$ as a function of $\kappa$, $\delta A(\tau_q)\propto \tau_q^{b}$ for different quantities $\hat{A}$ with $g_f=1$, $\omega\tau_q\in[10^3,10^{4}]$ and a $T=0$ bath. The universal AKZ predictions are indicated by dashed lines (see main text). Panel (d) shows the fitted exponents $b$ when $g_f=0.75$, which follow $\delta A(\tau_q)\sim\tau_q$. }}
\label{fig2}
\end{figure}

{\em Example.---} We illustrate the universal AKZ scaling laws in two $d=0$ systems which exhibit a mean-field QPT, namely, the QRM~\cite{Rabi:36} and the LMG model~\cite{Lipkin:65}. Models involving a mean-field QPT are of significance in diverse experimental platforms~\cite{Mottl:12,Brennecke:13,Anquez:16,Zibold:10,Jurcevic:17,Zhang:17b}. Although the QRM comprises two degrees of freedom, a spin and a single bosonic mode, it is possible to find a QPT in a suitable parameter limit~\cite{Hwang:15,Puebla:16,Puebla:17,sup}. In contrast, the LMG comprises $N$ two-level systems with a long-range interaction~\cite{Lipkin:65,Ribeiro:08,Ribeiro:07,Dusuel:04,Dusuel:05}. In the thermodynamic limit, denoted here by $\eta\rightarrow \infty$, the Hamiltonian of both models in one of the phases, $0\leq g\leq g_c=1$, can be written as~\cite{sup}
\begin{align}\label{eq:H0}
\hat{H}^{(0)}(g)=\omega \hat{a}^{\dagger}\hat{a} -\frac{g^2\omega}{4}(\hat{a}+\hat{a}^{\dagger})^2,
  \end{align}
where $\hat{a}$ and $\hat{a}^\dagger$ stand for the bosonic mode in the QRM and for the Holstein-Primakoff transformed pseudo-angular momenta in the LMG, and with $\omega$ its energy scale ($\hbar=1$). For the purposes of this Letter it is enough to consider one phase since KZ scaling laws appear when $g(\tau_q)=g_c$~\cite{Hwang:15,Defenu:18}, and so we consider $g(t)=g_ft/\tau_q$ with $g_f\leq g_c$. The Eq.~\eqref{eq:H0} describes the low-energy subspace, where we find that the eigenstates are $\ket{\phi_k(g)}=\hat{\mathcal{S}}[s(g)]\ket{k}$ with $\ket{k}$ the $k$th eigenstate of $\hat{a}^{\dagger}\hat{a}$, $\hat{\mathcal{S}}[s]=e^{\frac{1}{2}(s^*a^{2}-sa^{\dagger,2})}$ and $s(g)=\frac{1}{4}\ln(1-g^{2})$, so that $\hat{H}^{(0)}(g)=\sum_k \epsilon_k(g) \ket{\phi_k(g)}\bra{\phi_k(g)}$ and $\epsilon_k(g)=k\omega\sqrt{1-g^{2}}$. The low-energy excited states inherit thus the critical properties of the QPT. In particular, the number of bosons diverges $\langle \hat{a}^{\dagger}\hat{a} \rangle_k\sim |g-g_c|^{-1/2}$ for $\ket{\phi_k(g)}$,  while the position and momentum quadrature become $\Delta x_k\sim |g-g_c|^{-1/4}$ and $\Delta p_k\sim |g-g_c|^{1/4}$, respectively, such that $\Delta x_k \Delta p_k = (2k+1)$. Hence, $\gamma_{\Delta p}=-\gamma_{\Delta x}=\gamma_{\adaga}/2=1/4$.  In addition, we compute the residual energy $E_r(\tau_q)\equiv {\rm Tr}[\hat{\rho}(\tau_q)\hat{H}^{(0)}(g(\tau_q))]-E_{\rm gs}(g(\tau_q))$ whose critical exponent is given by $\gamma_{E_r}=z\nu$ as it is related to the energy gap $\Delta_1\sim |g-g_c|^{z\nu}$, with $z\nu=1/2$. Note that the energy gap for the $k$th eigenstate is $\Delta_k(g)=\epsilon_k(g)-\epsilon_0(g)=k\omega\sqrt{1-g^2}$ and thus universal AKZ scaling for $\delta E_r$ is expected too. Substituting $\gamma_A$ for these quantities in Eq.~\eqref{eq:deltaAKZ},  we obtain their predicted AKZ scaling, namely, $\delta \hat{a}^{\dagger}\hat{a}(\tau_q)\sim \tau_q^{4/3}$,  $\delta\Delta x\sim \tau_q^{7/6}$, $\delta \Delta p\sim \tau_q^{5/6}$ and $\delta E_r\sim \tau_q^{2/3}$ provided $g(\tau_q)=g_c$. The universal AKZ scaling breaks down if $g_f<g_c$, recovering the linear relation $\delta A(\tau_q)\sim \tau_q$.

In order to verify the AKZ scaling relations, we consider the system interacting weakly with a \emph{Markovian} bath at temperature $T$, although similar results are found considering a more realistic scenario~\cite{sup}, including possible memory effects via the non-perturbative approach developed in~\cite{Tamascelli:18,Mascherpa:19}. Thus, consider for now the dynamics fixed by the master equation~\cite{Breuer,Rivas}
\begin{align}\label{eq:ME}
\dot{\hat{\rho}}(t)=-i[\hat{H}^{(0)}(g(t)),\hat{\rho}(t)]+\mathcal{D}_{\hat{a}}[\hat{\rho}(t)]+\mathcal{D}_{\hat{a}^{\dagger}}[\hat{\rho}(t)],
  \end{align}
where $\mathcal{D}_{\hat{o}}[\bullet]=\Gamma_{\hat{o}}(2\hat{o}\bullet\hat{o}^{\dagger}-\{\hat{o}^{\dagger}\hat{o},\bullet \})$ is the Lindblad operator associated to the jump operator $\hat{o}$ and with rates $\Gamma_{\hat{a}}=\kappa(N_{\rm th}+1)/2$ and $\Gamma_{\hat{a}^{\dagger}}=\kappa N_{\rm th}/2$, with $N_{\rm th}=(e^{\omega/k_BT}-1)^{-1}$ the number of thermal excitations at temperature $T$. As the initial state is considered to be the ground state of $\hat{H}^{(0)}(0)$, we exploit the Gaussian-preserving nature of Eq.~\eqref{eq:ME}. For that, we employ the Wigner characteristic function $\chi(\beta,\beta^*,t)={\rm Tr}[e^{\beta \hat{a}^{\dagger}-\beta^*\hat{a}}\hat{\rho}(t)]$ with $\beta,\beta^*\in \mathbb{C}$ to calculate the evolution. The Fokker-Planck equation $\dot{\chi}(\beta,\beta^*,t)=\mathcal{X}[\chi(\beta,\beta^*,t)]$  allows for a Gaussian Ansatz, $\chi(\beta,\beta^*,t)=e^{i{\bf u}^t\mu(t)-\frac{1}{2}{\bf u}^t\Sigma(t){\bf u}}$ with ${\bf u}=(\beta,\beta^*)^t$, and first and second moments $\mu(t)=(q_0(t),q_1(t))^t$ and $\Sigma(t)=(\sigma_{00}(t),\sigma_{01}(t); \sigma_{10}(t), \sigma_{11}(t))$, respectively. Defining $2\sigma(t)=\sigma_{00}(t)+\sigma_{11}(t)$, and since $q_{0,1}(0)=0$, we obtain~\cite{sup}
\begin{align}\label{eq:FP1}
  \dot{\sigma}(t)&=2\Gamma_{-}\sigma(t)+\Gamma_++iG(t)(\sigma_{01}(t)-\sigma_{10}(t))\\ \label{eq:FP2} 
  \dot{\sigma}_{10}(t)&=2(i\omega-iG(t)+\Gamma_{-})\sigma_{10}(t)+i2G(t)\omega\sigma(t)
  \end{align}
with $\sigma_{01}(t)=\sigma_{10}^*(t)$, $G(t)=g^2(t)\omega/2$, $\Gamma_{\pm}=\Gamma_{\hat{a}^{\dagger}}\pm \Gamma_{\hat{a}}$, $\sigma(0)=1/2$ and $\sigma_{01}(0)=0$. From $\chi(\beta,\beta^*,t)$ we calculate the quantities of interest, e.g. ${\rm Tr}[\hat{a}^{\dagger}\hat{a}\hat{\rho}(t)]=\sigma(t)-\frac{1}{2}$ and ${\rm Tr}[(\hat{a}+\hat{a}^{\dagger})^2\hat{\rho}(t)]=2\sigma(t)-\sigma_{01}(t)-\sigma_{10}(t)$~\cite{sup}. Solving Eqs.~\eqref{eq:FP1}-\eqref{eq:FP2} under $g(t)$ with different quench times $\tau_q$ allows us to obtain $\langle \hat{A}(\tau_q)\rangle$ for the isolated case ($\kappa=0$) and $\langle \hat{A}(\tau_q)\rangle_{\rm op}$ for a chosen $\kappa\neq 0$ and $T$, with $\hat{A}\in \{\hat{a}^{\dagger}\hat{a},\Delta x,\Delta p,E_r \}$. Then, we calculate the excess due to the dissipation as $\delta\hat{A}(\tau_q)=\langle \hat{A}(\tau_q)\rangle_{\rm op}-\langle \hat{A}(\tau_q)\rangle$  to corroborate the AKZ scaling prediction (cf. Eq.~\eqref{eq:deltaAKZ}).

In Fig.~\ref{fig2}(a) we show the results for $\langle E_r(\tau_q)\rangle_{\rm op}$ together with $\langle E_r(\tau_q)\rangle$ for $\kappa=10^{-4}\omega$ and $T=10\omega$ when the quench, $g(t)=g_ft/\tau_q$, ends either at $g_f=0.75$ or at the critical point $g_f=1$. As expected, $\langle E_r(\tau_q)\rangle$ follows the KZ prediction $\tau_q^{-z\nu/(z\nu+1)}=\tau_q^{-1/3}$ as $z\nu=1/2$ when $g_f=1$, and the adiabatic scaling $\tau_q^{-2}$ for $g_f<1$~\cite{Hwang:15}. For $\kappa\neq 0$, $\langle E_r(\tau_q)\rangle_{\rm op}$ deviates from its isolated value. In Fig.~\ref{fig2}(b) we show the excess $\delta E_r(\tau_q)$: for $g_f<1$ it follows a linear scaling $\delta E_r(\tau_q)\sim \tau_q$, while for $g_f=1$ it shows the universal AKZ scaling given in Eq.~\eqref{eq:deltaAKZ}, $\delta E_r(\tau_q)\sim \tau_q^{(z\nu+1-\gamma_{E_r})/(z\nu+1)}=\tau_q^{2/3}$. Note that for $\tau_q\kappa\gtrsim 1$, $\langle E_r(\tau_q)\rangle_{\rm op}$ saturates to a constant value and so does $\delta E_r(\tau_q)$, since  $\tau_q$ is long enough to reach a steady state. Similar behavior is observed for other quantities. We compare the numerically determined exponent $b$ from a fit to $\delta A(\tau_q)=a\tau_q^b$ with the AKZ scaling prediction $(z\nu+1-\gamma_{A})/(z\nu+1)$ as a function of $\kappa$ for $\hat{A}\in \{\hat{a}^{\dagger}\hat{a},\Delta x,\Delta p,E_r \}$ in the interval $\omega\tau_q\in [10^{3},10^{4}]$ and $g_f=1$. This is plotted in Fig.~\ref{fig2}(c) choosing a $T=0$ bath, while Fig.~\ref{fig2}(d) shows the results when $g_f=0.75$. The latter illustrates how the universal AKZ is lost in favor of the linear scaling when the excited states are not critical. As for the standard KZ scaling, some quantities may exhibit stronger corrections to its leading scaling, as found here for $\delta \hat{a}^{\dagger}\hat{a}(\tau_q)$ and $\delta \Delta x (\tau_q)$~\cite{sup}. 

\begin{figure}
\centering
\includegraphics[width=1.\linewidth,angle=0]{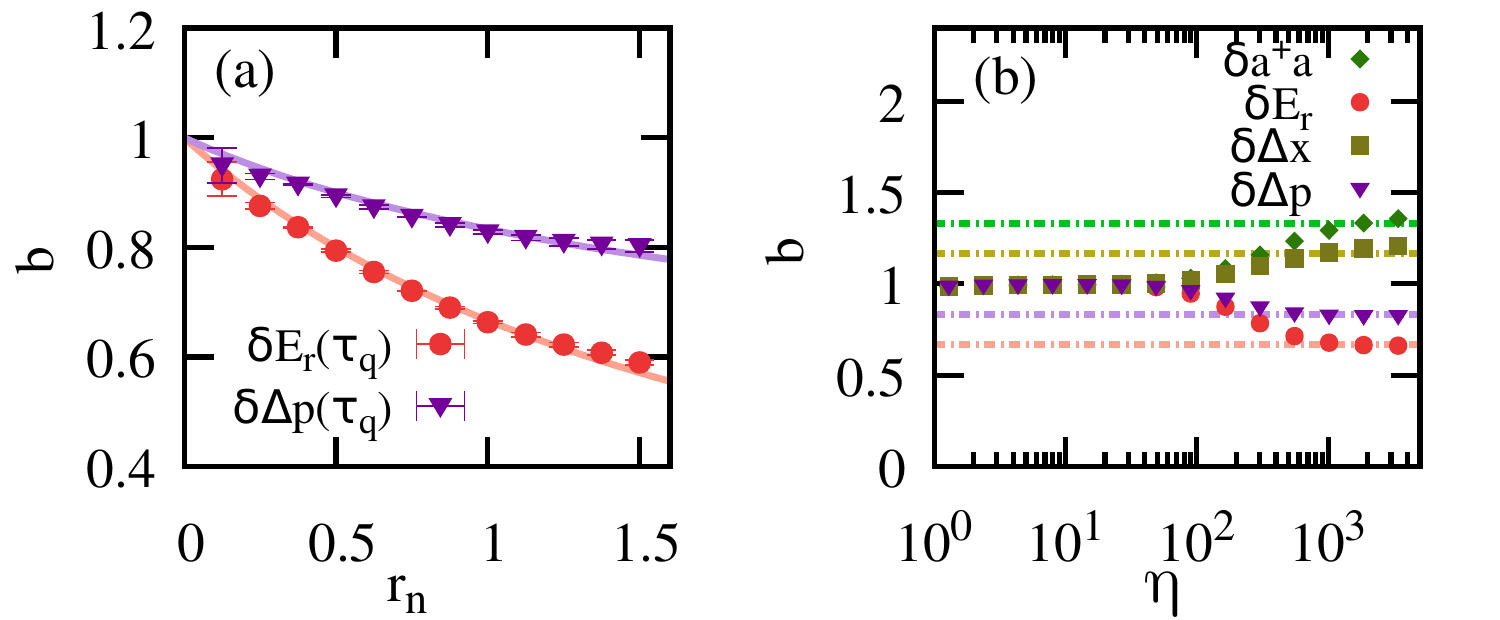}
\caption{\small{(a) Numerically determined AKZ scaling exponent $b$ under a non-linear ramp $g(t)\propto (t/\tau_q)^{r_n}$ as a function of $r_n$, for $\delta E_r(\tau_q)$ (red circles) and $\delta \Delta p(\tau_q)$ (magenta triangles) together with the universal AKZ prediction (solid lines), $2/(r_n+2)$ and $(r_n+4)/(2 r_n+4)$, respectively. The fits performed in $\omega\tau_q\in[10^4,10^5]$ with $\kappa=10^{-6}\omega$ and $T=0$. In panel (b) we show the dependence of $b$ with the QRM system size $\eta$ for $\kappa=10^{-5}\omega$ and $\omega\tau_q\in[10^3,10^4]$.}}
\label{fig3}
\end{figure}

{\em Non-linear ramps.---} The scaling presented in Eq.~\eqref{eq:deltaAKZ} is valid for linear ramps. However, as for the standard KZ predictions~\cite{Sen:08,Barankov:08}, non-linear ramps $g(t)\propto (t/\tau_q)^{r_n}$ with an exponent $r_n>0$ non-trivially modify the universal AKZ scaling laws. Here we choose $g(t)=g_f(1-(1-t/\tau_q)^{r_n})$. The adiabatic-impulse boundary becomes $|\tilde{g}-g_c|\sim \tau_q^{-r_n/(z\nu r_n+1)}$ as it can be shown relying on adiabatic perturbation theory~\cite{sup}, or by scaling arguments~\cite{Sen:08,Barankov:08}. Then, provided $g(\tau_q)=1$ this non-linear ramp leads to~\cite{sup}
  \begin{align}\label{eq:deltaAKZ_NL}
\delta A(\tau_q)\sim \tau_q^{(z\nu r_n+1-\gamma_A r_n)/(z\nu r_n+1)}.
  \end{align}
  As an example we consider the scaling for $\delta E_r(\tau_q)$ and $\delta \Delta p$, which are expected to follow $\delta E_r(\tau_q)\sim \tau_q^{2/(r_n+2)}$ and $\delta \Delta p(\tau_q)\sim \tau_q^{(r_n+4)/(2r_n+4)}$. In Fig.~\ref{fig3}(a) we show the fitted exponent $b$ as a function of $r_n$ (see caption for parameters), which agree very well with the universal AKZ prediction. In particular, for $r_n=1/2$, we find $b=0.794(3)$ and $0.893(3)$ for $\delta E_r(\tau_q)$ and $\delta\Delta p$, close to the predicted values $4/5$ and $9/10$. Similar results are obtained for other quantities~\cite{sup}.


  {\em Finite-size effects.---} The previous results have been obtained in the  thermodynamic limit, $\eta\rightarrow \infty$.  Finite-size systems however do not feature true singularities, e.g., they possess finite energy gap at the critical point. This leads to deviations from KZ scaling laws and to a maximum quench time in order to apply KZ arguments.   It is therefore advisable to investigate how these AKZ scaling laws emerge as the system size increases. For that, we consider  $\hat{H}(g)=\hat{H}^{(0)}(g)+\eta^{-1}\hat{H}^{(1)}(g)+O(\eta^{-2})$ where $\hat{H}^{(0)}(g)$ is again the Hamiltonian for $\eta\rightarrow\infty$, Eq.~\eqref{eq:H0}. The first-order correction $\hat{H}^{(1)}(g)$ comprises up to quartic terms in $\hat{a}$ and $\hat{a}^{\dagger}$, although its specific form depends on the model, namely $\hat{H}^{(1)}_{\rm QRM}(g)=\frac{g^4\omega}{16}f_{\rm QRM}(\hat{a},\hat{a}^{\dagger})$, and $\hat{H}^{(1)}_{\rm LMG}(g)=\frac{g^2\omega}{8}f_{\rm LMG}(\hat{a},\hat{a}^{\dagger})$.
  In order to capture the main finite-size effects we keep only quadratic terms upon normal ordering in $\hat{H}^{(1)}_{\rm QRM,LMG}(g)$ that preserves the Gaussian form of $\chi(\beta,\beta^*,t)$ but evolving now under modified equations of motion. In particular, $f_{\rm QRM}(\hat{a},\hat{a}^{\dagger})\approx 12\hat{a}^{\dagger}\hat{a}+6(\hat{a}^2+\hat{a}^{\dagger,2})+3$, while $f_{\rm LMG}(\hat{a},\hat{a}^{\dagger})\approx 4\hat{a}^{\dagger}\hat{a}+\hat{a}^2+\hat{a}^{\dagger,2}$. For the QRM (the analogous results for the LMG are shown in~\cite{sup}), the equations of motion follow from Eqs.~\eqref{eq:FP1}-\eqref{eq:FP2} replacing $G(t)$ by $G_{\rm QRM}(t)=g^2(t)\omega/2-12g^4(t)\omega/\eta$. Hence, in the $\eta\rightarrow \infty$ limit,  Eqs.~\eqref{eq:FP1}-\eqref{eq:FP2} are recovered.
The results plotted in Fig.~\ref{fig3}(b) reveal a smooth crossover of the exponent $b$ from linear, $b=1$, to its universal AKZ scaling value as the system size increases ($\eta\gtrsim 10^3$). The system size at which the crossover takes place in this fully connected model depends on the chosen quench time interval~\cite{sup}.

{\em Conclusions.---}
We have shown that the system-environment interaction can lead to universal scaling laws  upon a ramp towards the critical point of a QPT, similar to KZ scaling laws for isolated systems. Even a weak system-environment interaction yields an AKZ scaling, i.e., slower ramps provoke more excitations. Remarkably, provided the excited states of the quantum system display singular behavior, the AKZ scaling acquires a universal form, i.e., relevant observables scale in a power-law fashion with the quench time, whose scaling is determined solely by the equilibrium critical exponents of the QPT. If the excited states do not display critical behavior, a linear AKZ scaling  is recovered~\cite{Dutta:16}. We illustrated these findings in fully-connected, i.e. zero-dimensional, interacting systems such as the QRM and LMG undergoing dissipation, which are examples of superradiant~\cite{Hwang:15,Larson:17,Peng:19} and ferromagnetic QPT, respectively.  The results are also extended to non-linear ramps, and we further showed how the universal AKZ scaling laws emerge increasing the system size. 
The reported results may stimulate further research to underpin the role of open quantum dynamics in critical phenomena, and how its universal traits are modified when including a non-negligible interaction with a bosonic or fermionic environment of local or non-local nature, with distinct coupling directions~\cite{Arceci:17}, and systems with different spatial dimensions.

\begin{acknowledgments}
The authors thank Mauro Paternostro for his careful reading of the manuscript and valuable comments. R. P. acknowledges the support by the SFI-DfE Investigator Programme (grant 15/IA/2864). A. S., M. B. P. and S. F. H. acknowledge support of the ERC Synergy grant BioQ.
\end{acknowledgments}


%

\widetext
\clearpage
\begin{center}
\textbf{\large Supplemental Material\\Universal anti-Kibble-Zurek scaling in fully-connected systems}
\end{center}
\setcounter{equation}{0}
\setcounter{figure}{0}
\setcounter{table}{0}
\setcounter{page}{1}
\makeatletter
\renewcommand{\theequation}{S\arabic{equation}}
\renewcommand{\thefigure}{S\arabic{figure}}
\renewcommand{\bibnumfmt}[1]{[S#1]}
\renewcommand{\citenumfont}[1]{S#1}

\begin{center}
  Ricardo Puebla,${}^1$ Andrea Smirne,${}^{2,3}$ Susana F. Huelga,${}^2$ Martin B. Plenio${}^2$\\
\vspace{0.2cm}{\small $^1${\em Centre for Theoretical Atomic, Molecular, and Optical Physics,\\ School of Mathematics and Physics, Queen's University Belfast, Belfast BT7 1NN, United Kingdom}\\
  $^2${\em Institute of Theoretical Physics and IQST, Albert-Einstein Allee 11, Universit\"{a}t Ulm,
        89069 Ulm, Germany}
  \\
  $^3${\em Dipartimento di Fisica ''Aldo Pontremoli'', Universit\`a degli Studi di Milano, \\e Istituto Nazionale di Fisica Nucleare, Sezione di Milano, Via Celoria 16, I-20133 Milan, Italy}}
  \end{center}

\section{Further details on Eq. (3) of main text}
Here we provide more details on 
the validity of the derived universal anti-Kibble-Zurek scaling laws, given in Eq. (3) of the main text, which is reproduced here for convenience:
\begin{align}\label{SM_eq:uAKZ}
\delta A(\tau_q)\sim \tau_q^{(z\nu+1-\gamma_A)/(z\nu+1)},
\end{align}
where $z\nu$ and $\gamma_A$ are the equilibrium critical exponents of the energy gap and the observable $\hat{A}$. We recall that such an expression is obtained from
\begin{align}\label{SM_eq:deltaA}
\delta A(\tau_q)\approx \kappa \tau_q \sum_{k,l=0}h_{k,l} \langle\hat{A}\rangle_{k,l}, 
  \end{align}
which is Eq. (2) in the main text; the latter accounts for the difference among the isolated case, $\langle \hat{A}(\tau_q)\rangle$, and the value upon a non-unitary time evolution, denoted by $\langle \hat{A}(\tau_q)\rangle_{\rm op}\equiv {\rm Tr}[\hat{\rho}(\tau_q)\hat{A}]$, so that $\delta A(\tau_q)={\rm Tr}[\hat{\rho}(\tau_q)\hat{A}]-\langle \hat{A}(\tau_q)\rangle$. In the following derivation we assume a linear quench $g(t)\propto t/\tau_q$ and a final value  $g_f\equiv g(\tau_q)$. In addition, recall that the coefficients $h_{k,l}$ describe how the dissipative dynamics populates different excited states and their coherences, while  $\langle\hat{A}\rangle_{k,l}\equiv \langle \phi_k(g_f)|\hat{A} |\phi_l(g_f)\rangle $ denotes the matrix element of the operator $\hat{A}$ between the $k$th and $l$th eigenstates of the Hamiltonian
at the final time $\tau_q$, so that $\hat{H}_S(g(\tau_q))=\hat{H}_S(g_f)=\sum_{k=0}\epsilon_{k}|\phi_k(g_f)\rangle \langle \phi_k(g_f)|$.  In particular, for the non-unitary evolution,
we consider $\dot{\hat{\rho}}(t)=\mathcal{L}(t)\hat{\rho}(t)$, with $\mathcal{L}(t)\hat{\rho}=-i[\hat{H}_S(t),\hat{\rho}]+\mathcal{D}[\rho]$ where $\mathcal{D}[\hat{\rho}]$ denotes the dissipator acting on the state $\hat{\rho}$ and it is time-independent, i.e., it has the form of a Lindblad dissipator~\cite{SM_Breuer,SM_Rivas};
moreover, it is proportional to an overall rate $\kappa$, which accounts for the strength of the interaction with the environment. From the master equation, we can formally write $\hat{\rho}(\tau_q)=\mathcal{T}e^{\int_0^{\tau_q}dt \mathcal{L}(t)}\hat{\rho}(0)=\sum_{k,l=0}\tilde{c}_{k,l} |\phi_l(g_f)\rangle \langle \phi_k(g_f)|$ while for the isolated case, $| \psi(\tau_q)\rangle=\mathcal{T}e^{-i\int_0^{\tau_q}dt \hat{H}_S(t)}\ket{\psi(0)}=\sum_{k=0} c_k |\phi_k(g_f)\rangle$, where $\mathcal{T}$ stands for the time ordering and the coefficients $c_k$ and $\tilde{c_k}$ fulfill $\sum_{k=0}|c_k|^2=\sum_{k=0}|\tilde{c}_{k,k}|^2=1$ and $c_{k,l}^*=c_{l,k}$, and although not written explicitly, they do depend on the total quench time $\tau_q$. Note that $\hat{\rho}(0)=\ket{\psi(0)}\bra{\psi(0)}$. In this manner, it follows that
\begin{align}
\delta A(\tau_q)={\rm Tr}[\hat{\rho}(\tau_q)\hat{A}]-\langle \hat{A}(\tau_q)\rangle=\sum_{k,l=0} (\tilde{c}_{k,l}-c_k^*c_l) \langle \phi_k(g_f)|\hat{A} |\phi_l(g_f)\rangle.
\end{align}
We may write the coefficients $\tilde{c}_{k,l}$ as $\tilde{c}_{k,l}= c_k^*c_l+v_{k,l}$ where $v_{k,l}$ accounts for the difference with respect to the unitary evolution. Note that $v_{k,l}$ depends on the total time of the quench $\tau_q$. Assuming a weak coupling, $\kappa\tau_q\ll 1$, we expand $v_{k,l}\approx \kappa \tau_q h_{k,l}$ where $h_{k,l}$ is now a constant factor independent on $\tau_q$ at leading order. In other words, we assume the dissipation as a weak perturbation in the system's evolution. In this manner, one finally achieves Eq.~\eqref{SM_eq:deltaA} (Eq. (2) of the main text), $\delta A(\tau_q)\approx \kappa \tau_q \sum_{k,l=0}h_{k,l} \langle\hat{A}\rangle_{k,l}$. 

The Eq.~\eqref{SM_eq:deltaA} dictates in general a linear scaling of $\delta A(\tau_q)$ as a function of the total quench time $\tau_q$, i.e. $\delta A(\tau_q)\sim \tau_q$, for $\kappa\tau_q\ll 1$. This is independent on the actual values of $h_{k,l}$ and $\langle\hat{A}\rangle_{k,l}$, provided that the latter does not feature singular behavior.

In the same manner than for standard adiabatic-impulse approximation (cf. Sec.~\ref{s:nonlin}), if $\langle\hat{A}\rangle_{k,l}\sim |g-g_c|^{\gamma_A}$, then 
\begin{align}
\delta A(\tau_q)\approx \kappa \tau_q \sum_{k,l=0}h_{k,l} \langle\hat{A}\rangle_{k,l}\sim \kappa\tau_q\sum_{k,l=0}h_{k,l}|g_f-g_c|^{\gamma_A}\sim \kappa \tau_q \sum_{k,l=0} h_{k,l} |\tilde{g}-g_c|^{\gamma_A}\sim \kappa \tau_q^{(z\nu+1-\gamma_A)/(z\nu+1)},
  \end{align}
where for the last part we have introduced the scaling of the adiabatic-impulse boundary $|\tilde{g}-g_c|\sim \tau_q^{-1/(z\nu+1)}$ and neglected constant factors. The previous expression is precisely the Eq. (3) given in the main text. The extension of the previous expression to non-linear quenches is straightforward (cf. Sec.~\ref{s:nonlin}). 

The scaling $\delta A(\tau_q)\sim \tau_q^{(z\nu+1-\gamma_A)/(z\nu+1)}$ relies on critical excited states, $\langle\hat{A}\rangle_{k,l}\sim |g-g_c|^{\gamma_A}$. However, a few clarifications about the validity and implications of this scaling are in order.

First, we stress that not all of the excited states are required to exhibit singular behavior
for Eq.(3) of the main text to hold. This can be seen, for example, by demanding $h_{k,l}=0$ $\forall k,l\neq n$ and $h_{n,n}\neq 0$, where $n$ denotes a specific excited state, for which $\langle\hat{A}\rangle_{n,n}\sim |g-g_c|^{\gamma_A}$, while $\langle\hat{A}\rangle_{k,l\neq n}$ do not necessarily exhibit a critical behavior. In this manner, $\delta A(\tau_q)\sim \tau_q^{(z\nu+1-\gamma_A)/(z\nu+1)}$ still holds.  Moreover, even if $h_{k,l}\neq 0$ $\forall k,l\neq n$ and $\langle\hat{A}\rangle_{k,l\neq n}$ are not singular, then
\begin{align}
  \delta A(\tau_q)\sim \kappa \tau_q^{(z\nu+1-\gamma_A)/(z\nu+1)}+\kappa \tau_q,
\end{align}
which shows that the linear scaling becomes sub-leading in favor of the universal AKZ scaling provided $\gamma_A<0$. For $\gamma_A\geq 0$, $\delta A(\tau_q)$ may exhibit different scaling behavior depending on the specific details of the system, although $\delta A(\tau_q)\sim \kappa \tau_q$ will dominate eventually in the long-time limit, $\tau_q\rightarrow\infty$ while keeping $\kappa\tau_q\ll 1$. Hence, the reported universal AKZ scaling is relevant in systems beyond the mean-field universality class, where only some (possibly even only one) of
the excited states exhibit critical behavior. Indeed, this opens the possibility to find a universal AKZ scaling in models with $d>0$ dimensions; this will be the subject of future investigation.

Second, finite-size effects will unavoidably smooth the critical behavior of the system. Indeed, the relation $\langle\hat{A}\rangle_{k,l}\sim |g-g_c|^{\gamma_A}$ can only take place in the strict thermodynamic limit. Yet, as we show in the main text, the reported universal AKZ scaling laws emerge even in finite-size systems (cf. Fig. 3 of main text). 

Third, as proposed in Ref.~\cite{Dutta:16}, the number of excitations may be reduced at an optimal driving time, whose value may show a universal power-law scaling in terms of the noise rate. In Ref.~\cite{Dutta:16} the authors consider a linear AKZ scaling of the number of excitations $n_{\rm ex} \approx r_o \tau_q +r_c\tau_q^{-\beta}$, with $r_o$ and $r_c$ dimensionful prefactors for the open and closed scaling, respectively. As it can be seen from previous derivations, $r_o\propto \kappa$.  The previous expression leads to an optimal quench time $\tau_{\rm opt}\sim r_o^{-1/(1+\beta)}$, which follows from $\partial n_{\rm ex}/\partial \tau_q|_{\tau_{\rm opt}}=0$ and $\partial^2 n_{\rm ex}/\partial \tau_q^2|_{\tau_{\rm opt}}>0$,  where $\beta$ is the universal KZ scaling prediction. In our case, combining the universal KZ and AKZ scaling, we can write
\begin{align}\label{SM_eq:scalingAop}
\langle \hat{A}(\tau_q)\rangle_{\rm op}\approx r_c \tau_q^{-\gamma_A/(z\nu+1)}+r_o \tau_q^{(z\nu+1-\gamma_A)/(z\nu+1)}.
  \end{align}
Provided $0<\gamma_A<z\nu+1$, the driving time at which $\langle \hat{A}(\tau_q)\rangle_{\rm op}$ becomes minimum reads as
\begin{align}\label{SM_eq:tauopt}
\tau_{\rm opt}\approx \frac{r_c\gamma_A}{r_o(z\nu+1-\gamma_A)}.
  \end{align}
For $-z\nu-1<\gamma_A<0$, the Eq.~\eqref{SM_eq:scalingAop} features an inflection point revealing the crossover among the two universal scalings at a time 
\begin{align}
\tau_{\rm inf-p}\propto \frac{r_c(1+\gamma_A+z\nu)}{r_o/(1-\gamma_A+z\nu)},
\end{align}
while for $|\gamma_A|>z\nu+1$ one finds $\partial^2 \langle \hat{A}(\tau_q)\rangle_{\rm op}/\partial \tau_q^2\neq 0$ $\forall \tau_q>0$. Note that Eq.~\eqref{SM_eq:tauopt} dictates a remarkably different scaling with respect to that reported in Ref.~\cite{Dutta:16}. We stress that the linear scaling of $\tau_{\rm opt}$ and $\tau_{\rm inf-p}$ with 
$1/r_o$
is a direct and measurable consequence of the universal AKZ scaling derived here.


\section{Lipkin-Meshkov-Glick model}\label{s:LMG}
The Lipkin-Meshkov-Glick (LMG) model~\cite{SM_Lipkin:65,SM_Ribeiro:08,SM_Ribeiro:07,SM_Dusuel:04} is a collective spin model with an infinitely long-ranged interaction. The Hamiltonian of the model can be written as
\begin{align}
\hat{H}_{\rm LMG}(g)=-\frac{\omega}{2}\sum_i\hat{\sigma}_i^z-\frac{g^2\omega}{2N}\sum_{i\leq j}\hat{\sigma}_i^x\hat{\sigma}_j^x
  \end{align}
where the $\hat{\sigma}_{i}^{x,y,z}$ represent the spin-$\frac{1}{2}$ Pauli matrices of each of the $N$ interacting spins, with an energy scale $\omega$. The dimensionless parameter $g$ accounts for the relative strength of the ferromagnetic spin coupling. Introducing the $N$-spin representation, i.e., $\hat{J}_{x,y,z}$ such that $[\hat{J}_i,\hat{J}_j]=i\epsilon_{ijk}\hat{J}_k$, with $\hat{J}_{\alpha}=\sum_{i=1}^N\hat{\sigma}_i^\alpha/2$ for $\alpha\in \{x,y,z\}$ the Hamiltonian becomes
\begin{align}
  \hat{H}_{\rm LMG}(g)=-\omega \hat{J}_z-\frac{g^2\omega}{N}\hat{J}_x^2+\frac{g^2\omega}{4}.
\end{align}
In thermodynamic limit, which here corresponds to $\eta\rightarrow \infty$ with $\eta=N$, this model shows a QPT at $g_c=1$~\cite{SM_Dusuel:04,SM_Dusuel:05}.
As $\hat{J}^2$ commutes with $\hat{H}_{\rm LMG}(g)$, we constrain ourselves to the subspace of maximum angular momentum $J=N/2$. Making use of the Holstein-Primakoff transformation, $\hat{J}_z=J-\hat{a}^{\dagger}\hat{a}$, $\hat{J}_+=2J\sqrt{1-\hat{a}^{\dagger}\hat{a}/(2J)}\hat{a}$, and $\hat{J}_-=2J\hat{a}^{\dagger}\sqrt{1-\hat{a}^{\dagger}\hat{a}/(2J)}$, with $\hat{J}_x=(\hat{J}_++\hat{J}_-)/2$, the Hamiltonian becomes 
\begin{align}\label{SM_eq:HLMG}
  \hat{H}_{\rm LMG}^{(0)}(g)=\omega \hat{a}^{\dagger}\hat{a}-\frac{g^2\omega}{4}(\hat{a}+\hat{a}^\dagger)^2+\frac{\omega(g^2-2N)}{4},
\end{align}
upon taking the $N\rightarrow \infty$ limit, i.e. $\hat{J}_{+}\approx 2J\hat{a}$ and $\hat{J}_{-}\approx 2J\hat{a}^{\dagger}$. The previous Hamiltonian is valid for $0\leq g\leq 1$. For $g>1$, a rotation of $\hat{J}_{x,y,z}$ is required to properly set the quantization axis in the Holstein-Primakoff transformation.  The Eq.~\eqref{SM_eq:HLMG}, without considering the constant energy shift, corresponds to $\hat{H}^{(0)}(g)$ in the main text. 

The Eq.~\eqref{SM_eq:HLMG} describes one of the phases of the LMG in the thermodynamic limit. For a large, yet finite number of spins $N$, we can calculate its leading order correction in $N^{-1}$ (see~\cite{SM_Dusuel:04,SM_Dusuel:05} for a meticulous calculation). For that, we keep the corrections $1/N$ from the Holstein-Primakoff transformation such that $\hat{H}_{\rm LMG}(g)=\hat{H}^{(0)}_{\rm LMG}(g)+\frac{1}{N}\hat{H}_{\rm LMG}^{(1)}+O(N^{-2})$. In particular, within this phase the correction reads as
\begin{align}
\hat{H}_{\rm LMG}^{(1)}(g)=\frac{g^2\omega}{8}\left(4 \hat{a}^{\dagger}\hat{a}+ \hat{a}^2+\hat{a}^{\dagger,2}+2\hat{a}^{\dagger}\hat{a}\hat{a}\hat{a}+4\hat{a}^{\dagger}\hat{a}^{\dagger}\hat{a}\hat{a}+2\hat{a}^{\dagger}\hat{a}^{\dagger}\hat{a}^{\dagger}\hat{a}\right).
  \end{align}
Under a Gaussian Ansatz, the function in the main text for the first-order correction becomes $f_{\rm LMG}(\hat{a},\hat{a}^{\dagger})=(4 \hat{a}^{\dagger}\hat{a}+ \hat{a}^2+\hat{a}^{\dagger,2})$.  

\section{Quantum Rabi model}\label{s:QRM}
The quantum Rabi model (QRM) Hamiltonian can be written as
\begin{align}
\hat{H}_{\rm QRM} (\lambda)= \frac{\Omega}{2}\hat{\sigma}_z+\omega\hat{a}^{\dagger}\hat{a}-\lambda \hat{\sigma}_x(\hat{a}+\hat{a}^{\dagger}).
\end{align}
where $\Omega$ corresponds to the qubit frequency, while $\omega$ denotes the frequency of the single mode  and $\lambda$ the coupling constant. In the $\etaIn$ limit with $\eta= \Omega/\omega$, the QRM adopts a simple form that reveals a QPT at $g_c=1$~\cite{SM_Hwang:15}, where $g$ is a dimensionless coupling constant $g\equiv 2\lambda/\sqrt{\Omega\omega}=2\lambda/(\omega\sqrt{\eta})$. The Hamiltonian becomes $\hat{U}^{\dagger}\hat{H}_{\rm QRM}(g) \hat{U}=\hat{H}^{(0)}_{\rm QRM}(g)+\frac{1}{\eta}\hat{H}^{(1)}_{\rm QRM}(g)+O(\eta^{-2})$ with $\hat{U}=e^{\hat{S}}$ and $\hat{S}=i\lambda/\Omega(\hat{a}+\hat{a}^{\dagger})\hat{\sigma}_y$. Thus, the contribution given by $\hat{H}^{(1)}(g)$ and higher-order terms vanish in the $\etaIn$ limit. Therefore, $\lim_{\etaIn}\hat{U}^{\dagger}\hat{H}_{\rm QRM}(g) \hat{U}=\hat{H}^{(0)}_{\rm QRM}(g)$, which upon projection onto the low-energy spin subspace reads as
\begin{align}\label{SM_eq:Hnp}
\hat{H}^{(0)}_{\rm QRM}(g)=\omega\hat{a}^{\dagger}\hat{a}-\frac{g^2\omega}{4}(\hat{a}+\hat{a}^{\dagger})^2-\frac{\Omega}{2},
  \end{align}
valid only within the normal phase, $0\leq g\leq g_c=1$. As for the LMG model, neglecting the constant energy shift, the previous Hamiltonian becomes $\hat{H}^{(0)}(g)$ given in the main text. The first-order correction in $\eta^{-1}$ is given by~\cite{SM_Hwang:15}
\begin{align}
\hat{H}^{(1)}_{\rm QRM}(g)=\frac{g^4\omega}{16}(\hat{a}+\hat{a}^{\dagger})^4,
\end{align}
so that $f_{\rm QRM}(\hat{a},\hat{a}^{\dagger})=12\hat{a}^{\dagger}\hat{a}+6(\hat{a}^2+\hat{a}^{\dagger,2})+3$.

\section{Dynamics of the characteristic function}\label{s:ch}
As the master equation $\dot{\hat{\rho}}(t)=-i[\hat{H}^{(0)}(g),\hat{\rho}(t)]+\mathcal{D}_{\hat{a}}[\hat{\rho}(t)]+\mathcal{D}_{\hat{a}^{\dagger}}[\hat{\rho}(t)]$ is quadratic in $\hat{a}$ and $\hat{a}^{\dagger}$, we make use of a Gaussian Ansatz for the Wigner characteristic function $\chi(\beta,\beta^*,t)={\rm Tr}[e^{\beta \hat{a}^{\dagger}-\beta^*\hat{a}}\hat{\rho}(t)]$. In particular, we introduce the following Ansatz
\begin{align}
\chi(\beta,\beta^*,t)=e^{i{\bf u}^t\mu(t)-\frac{1}{2}{\bf u}^t \Sigma(t){\bf u}},
  \end{align}
with the coefficients vector ${\bf u}$, first $\mu(t)$ and second $\Sigma(t)$ moments given by
\begin{align}
  {\bf u}&=\begin{bmatrix} \beta \\ \beta^* \end{bmatrix},\\
  \mu(t)&=\begin{bmatrix} q_0(t) \\ q_1(t) \end{bmatrix},\\
  \Sigma(t)&=\begin{bmatrix} \sigma_{00}(t) & \sigma_{01}(t) \\ \sigma_{10}(t) & \sigma_{11}(t) \end{bmatrix}.
\end{align}
The master equation we are considering reads as
\begin{align}\label{SM_eq:npLindblad}
\dot{\hat{\rho}}(t)=-i[\hat{H}^{(0)}(g(t)),\hat{\rho}(t)]+\mathcal{D}_{\hat{a}}[\hat{\rho}(t)]+\mathcal{D}_{\hat{a}^{\dagger}}[\hat{\rho}(t)]
  \end{align}
with $\mathcal{D}_{\hat{o}}[\bullet]=\Gamma_{\hat{o}}\left(2\hat{o}\bullet \hat{o}^{\dagger}-\left\{\hat{o}^{\dagger}\hat{o},\bullet \right\}\right)$ the Lindblad superoperator for the jump operator $\hat{o}$ with noise rate $\Gamma_{\hat{o}}$~\cite{SM_Breuer,SM_Rivas}. The previous master equation results in a Fokker-Planck equation for the characteristic function, $\dot{\chi}(\beta,\beta^*,t)=\mathcal{X}[\chi(\beta,\beta^*,t)]$ with
\begin{align}
\mathcal{X}[\bullet]=\left\{a_1^*\beta^*\partial_{\beta^*}+a_1\beta\partial_{\beta}+a_2|\beta|^2+a_3+a_4\partial_{\beta}\partial_{\beta^*}+a_5^*\beta^*\partial_{\beta}+a_5\beta\partial_{\beta^*}\right\}\bullet
  \end{align}
where the parameters are given by $a_1=i\omega-i\frac{g^2(t)\omega}{2}-\Gamma_{\hat{a}}+\Gamma_{\hat{a}^{\dagger}}$, $a_2=-\Gamma_{\hat{a}}-\Gamma_{\hat{a}^{\dagger}}$, $a_3=a_4=0$ and $a_5=\frac{i}{2}g^2(t)\omega$. Since we consider as initial state the ground state of $\hat{H}^{(0)}(g)$ at $g(0)=0$, i.e. the vacuum, it follows $q_{0}(0)=q_{1}(0)=0$, $\sigma_{01}(0)=\sigma_{10}(0)=0$ and $\sigma(0)=1/2$ where $\sigma(t)=(\sigma_{00}(t)+\sigma_{11}(t))/2$. This brings us to the following equations of motion,
\begin{align}
\label{SM_eq:sFP1}
  \dot{\sigma}(t)&=2\Gamma_{-}\sigma(t)+\Gamma_++i\frac{g^2(t)\omega}{2}(\sigma_{01}(t)-\sigma_{10}(t))\\ \label{SM_eq:sFP2} 
  \dot{\sigma}_{10}(t)&=(2i\omega-ig^2(t)+2\Gamma_{-})\sigma_{10}(t)+ig^2(t)\omega\sigma(t)\\ \label{SM_eq:sFP3}
    \dot{\sigma}_{01}(t)&=(-2i\omega+ig^2(t)+2\Gamma_{-})\sigma_{01}(t)-ig^2(t)\omega\sigma(t),
  \end{align}
where we have defined $\Gamma_{\pm}=\Gamma_{\hat{a}^{\dagger}}\pm \Gamma_{\hat{a}}$, and $q_{0}(t)=q_{1}(t)=0\ \forall t$. These expressions correspond to the Eqs. (6)-(7) in the main text. Solving these equations allows us to obtain relevant quantities, as for example the average number of bosonic excitations,
\begin{align}
\langle\hat{a}^{\dagger}\hat{a} \rangle ={\rm Tr}[\hat{a}^{\dagger}\hat{a} \hat{\rho}(t)e^{\beta \hat{a}^{\dagger}-\beta^*\hat{a}}]\left.\right|_{\beta=\beta*=0}=\left(-\partial_\beta\partial_{\beta^*}-\frac{1}{2}\right)\chi(\beta,\beta^*,t)\left.\right|_{\beta=\beta*=0}=\left(\sigma(t)+q_0(t)q_1(t)-\frac{1}{2} \right).
  \end{align}
In a straightforward manner, we  obtain $\langle\hat{x}^2\rangle$, $\langle\hat{p}^2\rangle$, $\langle\hat{x}\rangle$ and $\langle\hat{p}\rangle$, with $\hat{x}=\hat{a}+\hat{a}^{\dagger}$ and $\hat{p}=i(\hat{a}^{\dagger}-\hat{a})$. Since $q_{0,1}(t)=0$, it follows $\langle\hat{x}\rangle=\langle\hat{p}\rangle=0$, while $\langle \hat{x}^2\rangle=2\sigma(t)-\sigma_{01}(t)-\sigma_{10}(t)$ and  $\langle \hat{p}^2\rangle=2\sigma(t)+\sigma_{01}(t)+\sigma_{10}(t)$, so that $\Delta x=\sqrt{\langle\hat{x}^2\rangle-\langle\hat{x}\rangle^2}=\sqrt{2\sigma(t)-\sigma_{01}(t)-\sigma_{10}(t)}$ and $\Delta p=\sqrt{ 2\sigma(t)+\sigma_{01}(t)+\sigma_{10}(t)}$. The energy of the state at time $t$ follows then from $\langle\hat{a}^{\dagger}\hat{a} \rangle$ and $\langle\hat{x}^2\rangle=\langle(\hat{a}+\hat{a}^{\dagger})^2 \rangle$, 
\begin{align}
E(t)={\rm Tr}[\hat{H}^{(0)}(g(t))\hat{\rho}(t)]=\omega\left(\sigma(t)-\frac{1}{2}\right)-\frac{g^2(t)\omega}{4}\left(2\sigma(t)-\sigma_{01}(t)-\sigma_{10}(t)\right),
  \end{align}
from which the residual energy follows simply by subtracting the ground-state energy given by $E_{\rm gs}(g)=\omega\left(\sqrt{1-g^2}-1\right)/2$, so that $E_r(\tau_q)=E(\tau_q)-E_{\rm gs}(g(\tau_q))$.

\begin{figure}
\centering
\includegraphics[width=0.7\linewidth,angle=0]{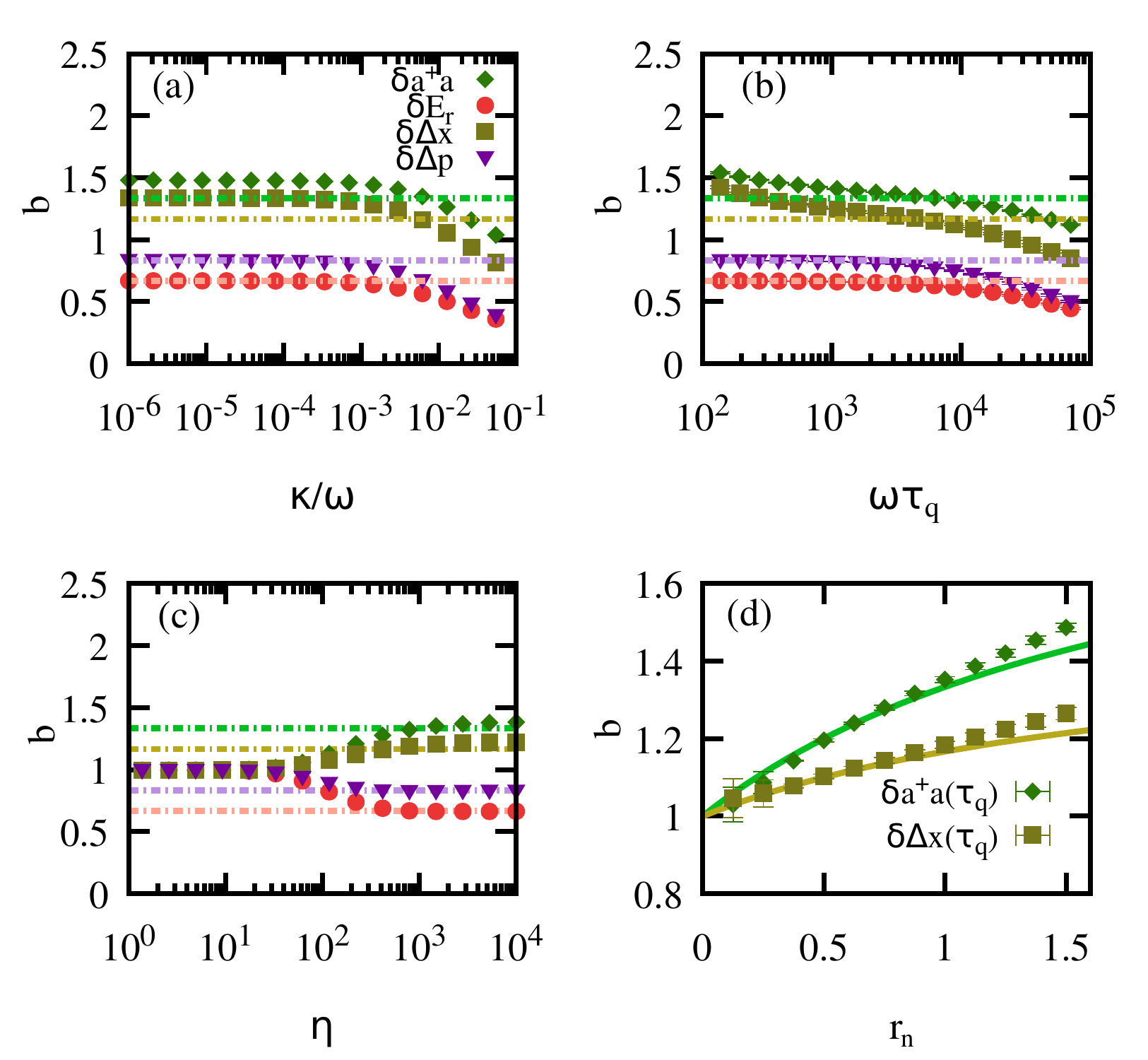}
\caption{\small{(a) Fitted scaling exponent $b$ for different quantities ($\delta\hat{a}^{\dagger}\hat{a}$, $\delta E_r$, $\delta \Delta x$ and $\delta \Delta p$)  for quench times $\omega\tau_q\in[10^2,10^3]$ with $g_f=1$ and $T=0$ and as a function of $\kappa$. The dashed lines correspond to the exponent predicted by the universal AKZ scaling. In panel (b) we show the dependence of $b$ on the chosen quench time $\omega\tau_q$ where the fit is performed. The panel (c) illustrates how the universal AKZ scaling emerges as the system size $\eta$ increases for a LMG model (recall that $\eta=N$ for the LMG). In the panel (d) we plot the resulting scaling exponent $b$ for $\delta\hat{a}^{\dagger}\hat{a}$ and $\delta \Delta x$ when performing non-linear ramps with exponent $r_n$. The fit was performed using $\kappa=10^{-6}\omega$ and for $\omega\tau_q\in[10^4,10^5]$. The lines show the predicted universal AKZ scaling under a protocol $g(t)=(1-(1-t/\tau_q)^{r_n})$, which for $\delta\hat{a}^{\dagger}\hat{a}$ and $\delta \Delta x$ results in $(2r_n+2)/(r_n+2)$ and $(3r_n+4)/(2r_n+4)$, respectively.}}
\label{fig_app}
\end{figure}

\section{Further results for the universal anti-Kibble-Zurek scaling laws}





As commented in the main text, the AKZ scaling for the quantities $\delta \hat{a}^{\dagger}\hat{a}$ and $\delta \Delta x$ tends to the predicted universal value for very long quench times $\tau_q$, yet requiring $\kappa\tau_q\ll 1$. For that, we solve the dynamics of the system and calculate these quantities, and then fit $\delta A(\tau_q)=a\tau_q^b$. The resulting exponent $b$ for $\delta \hat{a}^{\dagger}\hat{a}$ and $\delta \Delta x$ show a stronger dependence on the chosen interval $\tau_q$ where the fit is performed than for $\delta E_r$ and $\delta \Delta p$ (cf. Fig.~2(c) of the main text). In Fig.~\ref{fig_app}(a) we plot the numerically determined exponent $b$ for all these quantities (in the same format as in Fig.~2(c)) for the same parameters ($T=0$ and $g_f=1$) but performing the fit for shorter times, $\omega\tau_q\in[10^2,10^3]$. Note that the exponent $b$ for $\delta \hat{a}^{\dagger}\hat{a}$ and $\delta \Delta x$ remains larger than the predicted universal one, while for $\delta E_r$ and $\delta \Delta p$ we observe a very good agreement with the Eq. (3), given in the main text. This is due to stronger higher-order corrections. To investigate this, we have calculated the dependence of $b$ as a function of $\omega \tau_q$. This is plotted in Fig.~\ref{fig_app}(b) for $\kappa=10^{-4}\omega$, where one clearly observes the strong dependence of $b$ on $\tau_q$ for  $\delta \hat{a}^{\dagger}\hat{a}$ and $\delta \Delta x$, while at the same the scaling of $\delta E_r$ and $\delta \Delta p$ is more robust to the selected quench-time interval.


Any finite-size system with $\eta<\infty$ lifts the critical behavior found in the thermodynamics limit. However, as $\eta$ increases, the scaling laws are recovered, eventually leading into the predicted KZ and AKZ universal relations. In the main text we showed the behavior of the numerically fitted scaling exponents $b$ for the QRM, for which $\eta=\Omega/\omega$ as discussed in Sec. II and in the main text. As shown in Fig.~3(b) of the main text, the scaling exponent $b$ in the QRM for the different quantities considered here exhibit a crossover between $1$ (the standard AKZ scaling law due to the absence of critical behavior) and its universal value $(z\nu+1-\gamma_A)/(z\nu+1)$ as $\eta$ increases. Here we show that this is also the case for the LMG. In this case, $\hat{H}^{(1)}(g)=g^2\omega f_{\rm LMG}(\hat{a},\hat{a}^{\dagger})/8$ with $f_{\rm LMG}(\hat{a},\hat{a}^{\dagger})=4 \hat{a}^{\dagger}\hat{a}+ \hat{a}^2+\hat{a}^{\dagger,2}$ (see Sec. I). This leads into modified equations of motion for $\sigma(t)$, $\sigma_{10}(t)$ and $\sigma_{01}(t)$, which now depend on $\eta$ as
\begin{align}
\dot{\sigma}(t)&=2\Gamma_-\sigma(t)+\Gamma_++\frac{ig^2(t)\omega}{2}\left(1-\frac{1}{2\eta}\right)(\sigma_{01}(t)-\sigma_{10}(t))\\
\dot{\sigma}_{10}(t)&=\left[2i\omega+i\omega g^2(t)\left(\frac{1}{\eta}-1\right)+2\Gamma_-\right]\sigma_{01}(t)+i\omega g^2(t)\left(1-\frac{1}{2\eta}\right)\sigma(t),
  \end{align}
and $\sigma_{01}(t)=\sigma_{10}^*(t)$, with $\Gamma_{\pm}=\Gamma_{\hat{a}^\dagger}\pm \Gamma_{\hat{a}}$. Recall that for the LMG, $\eta=N$ is the number of spins. We solve the previous equations of motion and obtain the quantities $\delta \hat{a}^{\dagger}\hat{a}$, $\delta E_r$,  $\delta \Delta x$ and $\delta \Delta p$ as explained in Sec. III and in the main text. The scaling exponent for these quantities is plotted in Fig.~\ref{fig_app}(c) as a function of the system size $\eta$, where we have considered $T=0$ and $\kappa=10^{-5}\omega$, and the fit has been performed in the quench-time interval $\omega\tau_q\in[10^{3},10^{4}]$. The crossover is very similar to the one obtained for the QRM (cf. Fig.~3(b) of the main text). In addition, we remark that, depending on the chosen quench time interval where the fitted exponent $b$ is computed, the system size $\eta$ at which there is a crossover between linear ($b=1$) and the universal AKZ scaling ($b=(z\nu+1-\gamma_A)/(z\nu+1)$) changes.

In the main text we have shown the resulting AKZ scaling exponent for $\delta E_r$ and $\delta \Delta p$ when non-linear ramps are performed. In Fig.~\ref{fig_app}(d) we show the equivalent plot to Fig.~3(a) of the main text for $\delta \hat{a}^{\dagger}\hat{a}$ and $\delta \Delta x$. The universal AKZ scaling predicts $\delta \hat{a}^{\dagger}\hat{a}\sim \tau_q^{(2r_n+2)/(r_n+2)}$ and $\delta \Delta x\sim \tau_q^{(3r_n+4)/(2r_n+4)}$. The numerically fitted exponent $b$ (points) follows closely the prediction (lines). As for  the results plotted in Fig.~3(a), we have considered $T=0$ and $\kappa=10^{-6}\omega$, and the fit was performed in the interval $\omega\tau_q\in[10^4,10^5]$.

\section{Non-linear ramps: Scaling of the adiabatic-impulse boundary}\label{s:nonlin}
Here we show the modification introduced in the scaling of the adiabatic-impulse boundary when dealing with non-linear ramps $g(t)=g_f(1-(1-t/\tau_q)^{r_n})$, which yields $|\tilde{g}-g_c|\sim \tau_q^{-r_n/(z\nu r_n+1)}$ as commented in the main text (see also Refs.~\cite{SM_Sen:08,SM_Barankov:08}). For that, let us consider the time-evolved state under fully coherent dynamics, which can be expressed in the instantaneous eigenbasis of $\hat{H}^{(0)}(g(t))$, namely, $\ket{\psi(t)}=\sum_k \alpha_k(t) e^{-i\theta_k(t)}\ket{\phi_k(g(t))}$ with $\theta_k=\int_0^t ds \epsilon_k(g(s))$ the accumulated phase. We shall consider an initially prepared ground state, that is, $\alpha_0(0)=1$  and $\alpha_{k>0}(0)=0$. Then, the time evolution for $\alpha_k(t)$ follows from the Schr\"odinger equation,
\begin{align}
\alpha_k(t)=\alpha_k(0)-\int_0^t ds \sum_{l}\alpha_l(s) \langle \phi_k(g(s))|\partial_s|\phi_l(g(s))\rangle e^{i(\theta_k(s)-\theta_l(s))}.
  \end{align}
In the long $\omega\tau_q\gg 1$ limit, and since $\ket{\phi_k(g)}=\hat{\mathcal{S}}[s(g)]\ket{k}$  with $\hat{\mathcal{S}}[s]=e^{1/2(s^*\hat{a}^{2}-s\hat{a}^{\dagger,2})}$ and $s(g)=\ln(1-g^2)/4$ so that $\phi_k(g)|\partial_g|\phi_l(g)\rangle=\frac{1}{2}\frac{2g}{4(1-g^2)}(\sqrt{(k+2)(k+1)}\delta_{k+2,l}-\sqrt{k(k-1)}\delta_{k-2,l})$ one finds that the leading-order correction to the adiabatic case (i.e. the case in which $\alpha_0(t)=1$ $\forall t$) is given by
\begin{align}\label{eq:a2}
\alpha_2(g(t))\approx \frac{-ig_f r_n}{\tau_q}\frac{\sqrt{2}g(t)}{8\omega(1-g^2(t))^{z\nu+1}}\left(\frac{g_f}{g_f-g(t)} \right)^{(1-r_n)/r_n}e^{\frac{i\tau_q}{g_f r_n}\int_0^{g(t)}dg' \omega (1-g')^{z\nu}((g_f-g')/g_f)^{(1-r_n)/r_n}}.
  \end{align}
The boundary between adiabatic and impulse regimes can be estimated when the population in the excited state becomes of order $O(1)$, that is, from $|\alpha_2(\tilde{g})|^2\approx 1$. We remark that we are interested only in how $\tilde{g}$ scales with $\tau_q$. For that we set $g_f=g_c=1$ in Eq.~\eqref{eq:a2} and obtain $\tilde{g}$ from the previous condition, which yields the sought relation
\begin{align}\label{eq:gt_NL}
|\tilde{g}-g_c|\sim \tau_q^{-r_n/(z\nu r_n+1)}.
\end{align}

Having determined the scaling relation of the adiabatic-impulse boundary, Eq. (8) of the main text can be obtained as for the linear case. In particular, the excess in a quantity $\langle \hat{A}(\tau_q)\rangle_{\rm op}$ with respect to its isolated value, $\langle \hat{A}(\tau_q)\rangle$, scales as $\delta A(\tau_q)\sim \kappa \tau_q |\tilde{g}-g_c|^{\gamma_A}$ for sufficiently small $\kappa\tau_q$. Resorting to the adiabatic-impulse approximation when $g_f=g_c$, i.e., using Eq.~\eqref{eq:gt_NL}, one obtains
\begin{align}
  \delta A(\tau_q)\sim \kappa \tau_q \tau_q^{-r_n\gamma_A/(z\nu r_n+1)}\sim \tau_q^{(z\nu r_n+1-\gamma_A r_n)/(z\nu r_n+1)},
\end{align}
as given in the main text (cf. Eq. (8)).

\section{Results using optimized auxiliary oscillators}
Here we show the results supporting the universal AKZ scaling for a different, and more realistic type of system-environment interaction. For that, we rely on the method recently developed in~\cite{SM_Tamascelli:18,SM_Mascherpa:19},
exploiting that
the dynamics of a system interacting with a thermal environment is fixed by the spectral density $J(\omega)$ and the temperature $T$~\cite{SM_Breuer,SM_Rivas}, so that the dynamics can be effectively retrieved by introducing a number $N_a$ of interacting auxiliary oscillators with suitable parameters, which are obtained via a proper fitting procedure having $J(\omega)$ and $T$ as input. These auxiliary oscillators in turn are coupled to a zero-temperature local bath (see Ref.~\cite{SM_Mascherpa:19} for details about the method). We start considering the Hamiltonian of the system plus $N_a$ auxiliary oscillators $\hat{H}_{\rm SA}(g)=\hat{H}^{(0)}(g)+\hat{H}_{\rm I}+\hat{H}_{\rm A}$, with 
\begin{align}\label{SM_eq:HSA}
\hat{H}_{\rm SA}(g)= \omega\hat{a}^\dagger\hat{a}-\frac{g^2\omega}{4}(\hat{a}+\hat{a}^{\dagger})^2+\kappa (\hat{a}+\hat{a}^\dagger)\sum_{k=2}^{N_a+1} \left(c_k\hat{b}_k+c_k^{*}\hat{b}^{\dagger}_k\right)+\sum_{k=2}^{N_a+1} \omega_k \hat{b}_k^{\dagger}\hat{b}_k+\sum_{k=2}^{N_a+1} \left(d_k \hat{b}_k \hat{b}_{k+1}^{\dagger}+{\rm H.c.}\right),
  \end{align}
where $\kappa$ stands for a dimensionless coupling between the system and the environment, which is effectively described by the auxiliary oscillators, with operators $\hat{b}_k$ and $\hat{b}^{\dagger}_k$, and frequency $\omega_k$ for $k=2,\ldots, N_a+1$. Note that the oscillators interact through a hopping term between $k$ and $k+1$ with strength $d_k$. In addition, the auxiliary oscillators are locally damped, so that the dynamics of the system plus auxiliary oscillators is governed by the Lindblad equation
\begin{align}\label{SM_eq:rhoqrma}
\dot{\hat{\rho}}_{\rm SA}(t)=-i\left[\hat{H}_{\rm SA}(g(t)),\hat{\rho}_{\rm SA}(t)\right]+\sum_{k=2}^{N_a+1} \gamma_k\left( \hat{b}_k\hat{\rho}_{\rm SA}(t)\hat{b}_k^{\dagger}-\frac{1}{2}\left\{\hat{b}_k^{\dagger}\hat{b}_k,\hat{\rho}_{\rm SA}(t) \right\}\right).
\end{align}
The parameters of the model, i.e. the auxiliary oscillator frequencies $\omega_k$, couplings $d_k$ and $c_k$, as well as the rate $\gamma_k$, are fixed by $J(\omega)$ and $T$ via the mentioned fitting procedure. 
Importantly, the master equation given by Eq.~\eqref{SM_eq:rhoqrma} is of the Lindblad form, but it involves the auxiliary oscillators too. As a consequence, when restricting to the open-system degrees of freedom only, the resulting dynamics cannot be generally described by a Lindblad equation, contrary to what we considered in the main text [see Eq.~(5)]. The key point is in fact that Eq.~\eqref{SM_eq:HSA} allows us to keep track of possible memory effects into the evolution of the open system, thus yielding a more general and realistic description of the effects of the interaction with the environment.

As the initial state is a vacuum state, and the master equation is quadratic in the bosonic operators, we can employ Gaussian states theory. The Hamiltonian $\hat{H}^{(0)}_{\rm SA}$ can be written in a bilinear form as $\hat{H}^{(0)}_{\rm SA}(g)=\frac{1}{2}\hat{x}\cdot {\bf H}(g)\hat{x}$ with  $\hat{x}=\left(\hat{q}_1,\ldots,\hat{q}_{N_a+1},\hat{p}_1,\ldots,\hat{p}_{N_a+1}\right)^{T}$, with $\hat{a}=(\hat{q}_1+i\hat{p}_1)/\sqrt{2}$, $\hat{a}^{\dagger}=(\hat{q}_1-i\hat{p}_1)/\sqrt{2}$, $\hat{b}_k=(\hat{q}_{k}+i\hat{p}_{k})/\sqrt{2}$ and $\hat{b}^{\dagger}_k=(\hat{q}_{k}-i\hat{p}_{k})/\sqrt{2}$ with $k=2,\ldots,N_a+1$, and such that $[\hat{x}_j,\hat{x}_k]=i\hat{J}_{jk}$ and where $\hat{J}$ is the symplectic matrix
\begin{align}
  \hat{J}=\begin{bmatrix} {\bf 0}_{N_a+1} & {\bf I}_{N_a+1}\\
  -{\bf I}_{N_a+1} & {\bf 0}_{N_a+1}
  \end{bmatrix}. 
\end{align}
The dynamics can be cast in a time-dependent Lyapunov equation, 
\begin{align}\label{SM_eq:lyap}
\dot{{\bf V}}(t)={\bf V}(t){\bf \Gamma}^{T}(t)+{\bf \Gamma}(t){\bf V}(t)+{\bf D}
  \end{align}
with ${\bf V}_{jk}(t)=\langle\hat{x}_j\hat{x}_k+\hat{x}_k\hat{x}_j\rangle_t$, ${\bf \Gamma}(t)=\hat{J}\ {\bf H}(g(t))-{\rm Im}{\bf \Upsilon }\hat{J}$, ${\bf \Upsilon}=\sum_k\lambda_k\lambda_k^{\dagger}$ and ${\bf D}=2{\rm Re}{\bf \Upsilon}$. The vectors $\lambda_k$ are such that the Eq.~(\ref{SM_eq:rhoqrma}) can be written as $\dot{\hat{\rho}}_{\rm SA}(t)=-i[\hat{H}_{\rm SA}^{(0)},\hat{\rho}_{\rm SA}(t)]+\sum_k \left(\hat{L}_k \hat{\rho}_{\rm SA}(t) \hat{L}_k^{\dagger}-\frac{1}{2}\left\{\hat{L}_k^{\dagger}\hat{L}_k,\hat{\rho}_{\rm SA}(t) \right\}\right)$, with $\hat{L}_k=\lambda_k\cdot \hat{J} \hat{x}$, which contains the rates $\gamma_k$.
Since only the auxiliary oscillators undergo dissipation, $\hat{L}_1=0$. For the first auxiliary oscillator, $\hat{L}_2=\lambda_2\cdot \hat{J}\hat{x}=\sqrt{\gamma_2} \hat{b}_2=\frac{\sqrt{\gamma_2}}{\sqrt{2}}(\hat{q}_2+i\hat{p}_2)$, from where it immediately follows $\lambda_2$, and similarly for the rest. As the initial state is a vacuum state for all the oscillators, ${\bf V}(0)={\bf I}_{N_a+1}$. 

\begin{figure}
\centering
\includegraphics[width=0.7\linewidth,angle=0]{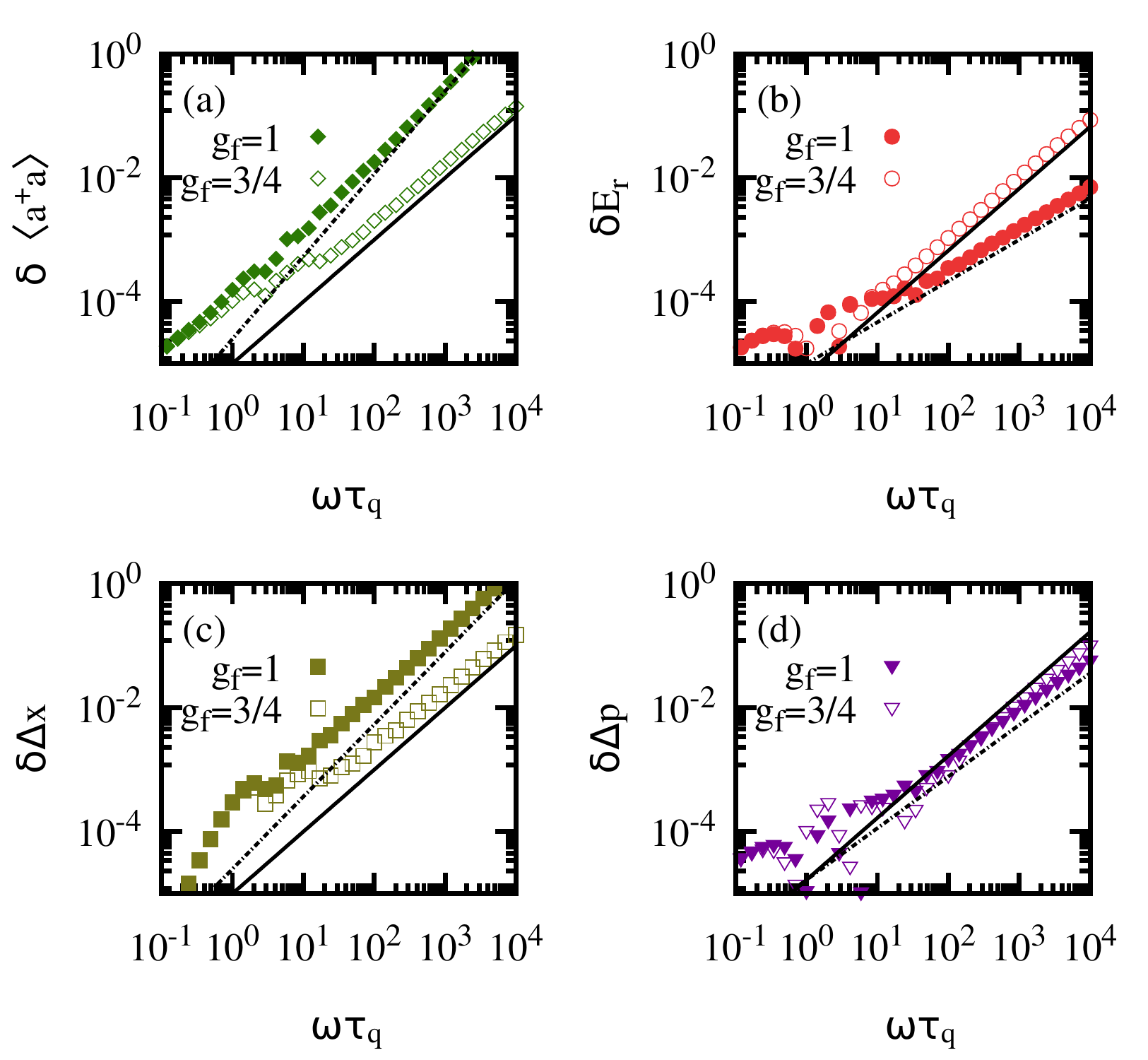}
\caption{\small{Scaling behavior of $\delta A$, for different quantities, namely, (a) $\delta\langle \hat{a}^\dagger\hat{a}\rangle$, (b) $\delta E_r$, (c) $\delta \Delta \hat{x}$ and (d) $\delta\Delta \hat{p}$, as a function of the quench time $\omega\tau_q$. The data corresponds to two different final couplings, namely, $g_f=1$ (critical point) and $g_f=3/4$, depicted with solid and open points, respectively. Other parameters are $\omega_c=20\omega$ and $\kappa=10^{-5}$. The predicted scaling is plotted with dashed and solid lines for the universal AKZ and linear scaling, respectively. Recall that the universal AKZ predicts $\delta A\propto \tau_q^{(z\nu+1-\gamma_A)/(z\nu+1)}$ with $\gamma_A$ the critical exponent associated to $A$. See Sec. VI for further details and the values of the numerically fitted exponents. }}
\label{SM_fig:OhmicKZ}
\end{figure}

Here we show that the predicted anti-Kibble-Zurek scaling laws hold for a system interacting with an environment characterized by an Ohmic bath at $T=0$, whose spectral density reads as
\begin{align}
J(\omega)=2\kappa^2\pi \omega e^{-\omega/\omega_c}.
  \end{align}
Then, employing the method described in~\cite{SM_Mascherpa:19}, its impact on the system can be effectively captured by $N=4$ auxiliary oscillators with the following parameters:
\begin{align}
\omega_2&=2.70796 \omega_c  \qquad d_2=3.38195\omega_c \qquad \gamma_2=11.9298\omega_c & \quad c_2&=(-0.0333215-0.0121362 i)\omega_c \\
\omega_3&=2.13014 \omega_c \qquad d_3=1.43514\omega_c\qquad \gamma_3=0.573494\omega_c & \quad c_3&=(0.319+0.0811955 i)\omega_c\\
\omega_4&=1.15884 \omega_c \qquad d_4=0.491546\omega_c \quad \ \ \gamma_4=0.0317143\omega_c & \quad c_4&=(0.760716+0.0175762 i)\omega_c\\
\omega_5&=0.310906 \omega_c \quad \ \  d_5=0 \qquad\qquad\quad \ \ \ \gamma_5=0.000795693\omega_c & \quad c_5&=(0.579218)\omega_c.
\end{align}

The dynamics of the system is obtained by numerically solving the Eq.~\eqref{SM_eq:lyap}, setting the previous parameters and with the time-dependent protocol $g(t)=g_f t/\tau_q$, as considered in the main text. Setting for example $\kappa=10^{-5}$ and $\omega_c=20\omega$ with $g_f=1$ and fitting the resulting $\delta A$ quantities to $\delta A\propto \tau_q^b$ in the range $\omega\tau_q\in[10^{2},10^{3}]$, we obtain $b=1.21(3)$, $0.66(3)$, $1.03(3)$ and $0.82(3)$ for $\delta\langle \hat{a}^\dagger\hat{a}\rangle$, $\delta E_r$, $\delta \Delta \hat{x}$ and $\delta\Delta \hat{p}$, respectively. Recall that the predicted anti-scaling exponents are $4/3$, $2/3$, $7/6$ and $5/6$, respectively the previous quantities. As in the case with a standard Lindblad master equation, we obtain again larger deviations with respect to the predicted AKZ values when $b>1$, i.e. for the exponents of $\delta\langle \hat{a}^\dagger\hat{a}\rangle$ and $\delta\Delta \hat{x}$ (cf. Sec. IV), while for $\delta E_r$ and $\delta\Delta\hat{p}$ the agreement between the numerically fitted exponent and the predicted ones is excellent. Similar results are obtained for different quench times provided $\kappa\omega\tau_q\ll 1$. In contrast, when the evolution does not end at the critical point, we expect a linear scaling $\delta A\propto \tau_q$. Choosing $g_f=3/4$, the numerically fitted exponents are $b=0.94(1)$, $0.98(1)$, $0.95(1)$ and $1.03(1)$, for $\delta\langle \hat{a}^\dagger\hat{a}\rangle$, $\delta E_r$, $\delta \Delta \hat{x}$ and $\delta\Delta \hat{p}$, respectively, and close to the predicted linear scaling. The results for these two cases ($g_f=1$ and $g_f=3/4$) are plotted in Fig.~\ref{SM_fig:OhmicKZ}. Finally, we also check the universal AKZ scaling exponents for a non-linear ramp. For that we choose a non-linear exponent $r_n=5/4$, and drive the system towards the critical point $g_f=1$, with the same parameters as before. For the same parameters, the numerically fitted scaling exponents are $b=1.30(1)$, $0.61(1)$, $1.10(1)$ and $0.79(1)$, close again to the predicted values in this case, namely, $18/13$, $8/13$, $31/26$ and $21/26$ (see main text for the predicted scaling exponents for non-linear ramps).
We leave as an open question for future investigations what happens in the presence of stronger couplings
and/or smaller cut-off frequencies $\omega_c$, i.e., when memory effects might be enhanced. 

%

\end{document}